# Critical roles of edge turbulent transport in the formation of high-field-side high-density front and density limit disruption in J-TEXT tokamak

Peng Shi[1,2,3], Yuhan Wang[2], Li Gao[2]*, Hongjuan Sun[1], Qinghu Yang[2], Xin Xu[2], Chengshuo Shen[2], Yanqiu Chen[2], Qinlin Tao[2], Zhipeng Chen[2], Haosheng Wu[4], Lu Wang[2], Zhongyong Chen[2], Nengchao Wang[2], Zhoujun Yang[2], Jingchun Li[5], Yonghua Ding[2], Yuan Pan[2] and J-TEXT team

[1]United Kingdom Atomic Energy Authority, Culham Centre for Fusion Energy, Culham Science Centre, Abingdon, Oxon, OX14 3DB, United Kingdom

[2]International Joint Research Laboratory of Magnetic Confinement Fusion and Plasma Physics, State Key Laboratory of Advanced Electromagnetic Engineering and Technology, School of Electrical and Electronic Engineering, Huazhong University of Science and Technology, Wuhan 430074, China

[3]Southwestern Institute of Physics, PO Box 432, Chengdu 610041, People's Republic of China

[4]NEMO Group, Dipartmento Energia, Politecnico di Torino, Corso Duca degli Abruzzi 24, 10129, Torino, Italy

[5]Department of Earth and Space Sciences, Southern University of Science and Technology, Shenzhen 518055, People's Republic of China

*Corresponding author: gaoli@mail.hust.edu.cn

**Abstract:**
This article presents an in-depth study of the sequence of events leading to density limit disruption in J-TEXT tokamak plasmas, with an emphasis on boudary turbulent transport and the high-field-side high-density (HFSHD) front. These phenomena were extensively investigated by using Langmuir probe and Polarimeter-interferometer diagnostics. The research reveals a consistent pattern of events as the plasma density ramps up: the collapse of the sheared radial electric field, the enhancement of a boundary broadband turbulence ($50{\sim}80 kHz$), the increase of boundary particle transport induced by this turbulence, edge cooling and the emergence of the HFSHD front. These phenomena occur once the plasma density exceeds a critical value. Importantly, by exploring plasmas with varying edge safety factor ($q_a$), it's revealed that the density thresholds for these phenomena are all inversely proportional to $q_a$. The findings offer valuable insights into the mechanisms underlying density limit disruptions in tokamak plasmas, suggesting that the enhancement of edge turbulent transport plays crucial roles in the edge cooling and triggering the HFSHD front. For the first time, a strong link between the edge turbulent transport and the HFSHD front has been observed. In addition, the evolution of boundary electron temperature across various $q_a$ plasmas can further validate the link between the edge turbulent transport and the HFSHD formation.

## 1. Introduction

High density is a prerequisite for the operation of future fusion reactors, since the power produced by fusion is proportional to the square of plasma density. To achieve higher economic efficiency in future large tokamaks, a high plasma density is highly desirable for the next-step tokamaks, such as the International Thermonuclear Experimental Reactor (ITER) and Chinese Fusion Engineering Testing Reactor (CFETR) [1-2]. However, experiments in tokamaks show that the plasma operating densities always have an upper limit. At present, the Greenwald density (line-averaged) scaling $n_G[10^{20}m^{-3}] = I_P[MA]/\pi a^2[m^2]$ is empirical for the density limit that is widely applicable to devices such as the Alcator, DIII and PBX [3-4].

Experimental and theoretical studies of the density limit have been pursued for decades, leading to a multitude of explanations for the underlying physical mechanisms. In experiments where the plasma density is increased via continuous gas puffing, various instabilities have been observed as the plasma density gradually ramps up to the density limit disruption. Results across different devices have revealed common characteristics of these physical phenomena during the process of density limit disruption [5-6]. For instance, thermal-radiation instabilities [7-8], edge cooling [5,9], boundary constraint deterioration, and eventual MHD instabilities [10-11] are consistently observed during the process of density ramp-up. The highly reproducible nature of these phenomena and their occurrence at the plasma boundary suggest that the boundary plasma behaviors play an important role in the physical mechanism of density limited disruption [12]. Moreover, it was found that the Greenwald density limit could be exceeded by optimizing fueling techniques, which can increase the central plasma density while maintain the boundary plasma density. This discovery provides strong evidence that the evolution of the boundary parameters plays a crucial role in determing the density limit [13].

Multifaceted Asymmetric Radiation From the Edge (MARFE), one of the most remarkable macro-phenomena observed in the process of density limit disruption, is widely believed to trigger the MHD instability and major disruption. MARFE is characterized by high electron density, low temperature, strong volumetric recombination, and commonly occurs at the high field side (HFS) near the inner wall of the tokamak in limiter [14]. Following the appearance of MARFE, the line integral density at the HFS boundary increases sharply and the radiated power in the MARFE region reaches 30-50% of the total heating power [14-15]. In divertor plasmas, the phenomena tend to manifest close to the X-point region, and often accompanied by divertor detachment, which is referred to as 'X-point MARFE' or 'divertor MARFE' [16-17].

In recent years, a high-field-side high-density (HFSHD) phenomenon, which is a poloidally localized high-density region in the HFS scrape-off layer (SOL), has been observed in the high-density operation on both divertor and limiter plasmas. In ASDEX-U and JET divertor plasmas, the HFSHD front appears near the inner target and moves to the X-point as the plasma density increases, followed by the X-point MARFE and divertor detachment [18-22]. In the J-TEXT limiter plasmas, the HFSHD front is found to form at the midplane of HFS edge, and tends to move towards the low field side (LFS) just prior to the density limit disruption [23]. The HFSHD front shares many similarities to MARFE, such as both occurring at the HFS boundaries and exhibiting localized high density. However, there are also distinct differences between the two phenomena. The radiation power from the region of HFSHD front is significantly lower than MARFE. And the HFSHD front is much stable at the inner target or HFS edge across a broad range of plasma density, which is in contrast to the unstable behavior of MARFE. According to the above observations, the author believe that the HFSHD front is an early form of MARFE.

In addition to thermal-radiation instabilities, micro-turbulence instability also plays an important role in boundary cooling and density limit disruption [24-25]. Previous experiments in J-TEXT have observed the collapse of edge flow shear and a decrease of the ratio of Reynolds power to turbulence production during the ramp-up of plasma density [5]. There weve also observations of increased electron particle and heat transport before density limit disruption [26-27]. These findings are consistent with observations on Alcator C-Mod [28] and HL-2A [29]. Furthermore, plasma confinement can be improved by maintaining the edge shear layer and suppressing the boundary particle flux, accomplished by applying a positive bias to [27,30]. This can confirm the roles of edge turbulent transport.

Among those factors potentially contributing to the density limit, multiple factors can interact simultaneously, and the intricate details of their interplay are still open questions. This paper aims to investigate the roles of boundary turbulence in determine density limit and its relationship between with HFSHD. Considering that the edge safety factor $q_a$ can potentially impact turbulence and transport, but has no significant effects on density limit, we have conducted a wide study of boundary turbulent transport and HFSHD front across plasmas with various $q_a$.

The rest of the paper is organized as follows: Section 2 describes the experimental setup and main diagnostics for the investigation of the edge turbulence behaviour and HFSHD front in the J-TEXT tokamak. Section 3 presents the experimental results during the density ramp-up. Section 4 presents conclusion and discussion.

## 2. Experimental Setup

The J-TEXT tokamak (formerly TEXT-U [31]) is a conventional medium-sized tokamak with a major radius of $R_0 = 1.05\ m$ and minor radius of $a = 0.25\sim0.29\ m$. The first wall and the limiter are covered with carbon tiles. In this experiment, we utilized three different positions of limiters - top, bottom and outer (Fig. 1(a)), all located at $r_{Limiter} = 25.5\ cm$. It's noteworthy that there is no limiter at the HFS in our experiments (it was removed recently for the HFS divertor operation) [32], which is quite different to the experiments in our previous publication [23]. The arrangement of the main diagnostics is shown in figure 1. The line-integrated electron density is measured by a multi-channel far-infrared laser polarimeter-interferometer system (POLARIS) [33]. This system views the plasma vertically at intervals of $3 cm$ in the radial mid-plane, ranging from $r = -24\ cm$ to $r = +24\ cm$, thereby covering the main plasma region ($|r| < 0.94a$), where $r = R - R_0$, as shown in Fig. 1(a) & (b). Here, $r < 0$ and $r > 0$ correspond to the HFS and the low field side (LFS), respectively.

A 36-channel photodiode array (PDA) system is used to measure the line emission of Hydrogen $\alpha$ ($H_\alpha$) [34]. This system, similar to POLARIS, also covers the HFS and LFS of the tokamak. The sight line of the inner-most three chords go through the area where the HFSHD is located, and only the channels employed in this experiment are showed in Fig. 1(a). The electrostatic probes arrays, also known as the combined Langmuir probe (CLP) [35], are mounted on the window Port#13 at the device's top (Fig. 1(b)). The configuration of the Langmuir probe array is shown in Fig. 1(c). The CLP consists of eight graphite probes, each with a diameter of 2 mm. Probes 1 - 4 are situated on step 1 with a length of 3 mm, while probes 5 - 8 are on step 2 with the same length. The connecting line of pins 5 and 7 aligns with the direction of the ring, avoiding the shadow effect between the probes. Pins 1 and 3 are distributed along the toroidal direction as a pair of Mach probes. Pins 2 and 4 measure the average floating potential on step1 $V_{f,step1} = (V_{f,2}+V_{f,4})/2$. The pins 6 and 8 spacing $d = 7mm$ are used to measure the average floating potential on step2 $V_{f,step2} = (V_{f,6}+V_{f,8})/2$. Pins 5 and 7 are biased to form a double probe for acquiring the ion saturation current $I_s = (V_{+,5} - V_{-,7})/R_{shunt}$, where $R_{shunt}$ refers to the sample resistor in the double probe circuit. Based on the above configuration, the electron temperature $T_e$, electron density $n_e$, the poloidal electric field $E_p$ and the radial electric field $E_r$ can be measured simutaneously by CLP. Electron temperature is inferred by $T_e = (V_{+,5} - V_{f,step2})/ln2$. Electron density is inferred by $n_e = I_s/(0.49eA_{eff}C_s)$, where $e$ is the elementary charge, $C_s = \sqrt{k_B(ZT_e + T_i)/m_i} \approx \sqrt{2k_BT_e/m_i}$ is ion sound speed and $A_{eff}$ is the effective current collection area. Plasma potential is inferred by $\varphi_p = V_{f,step2} + 2.5T_e$, and the radial electric field may be inferred from $E_r = -\nabla_r \varphi_p \approx -\nabla_r V_{f,step} - 2.5\nabla_r T_e/e$. When the electron temperature profile does not change significantly, the second term, $-2.5\nabla_r T_e/e$, is negligible. The poloidal electric field is computed as $E_p = (V_{f,6} - V_{f,8})/d$, here, $d$ is the spacing between pins 6 and 8. And the radial electric field $E_r$ can be esitmated by the gradient of floating potential in two steps as $E_r = -\nabla_r \varphi_p \approx -\nabla_r V_{f,step}$.

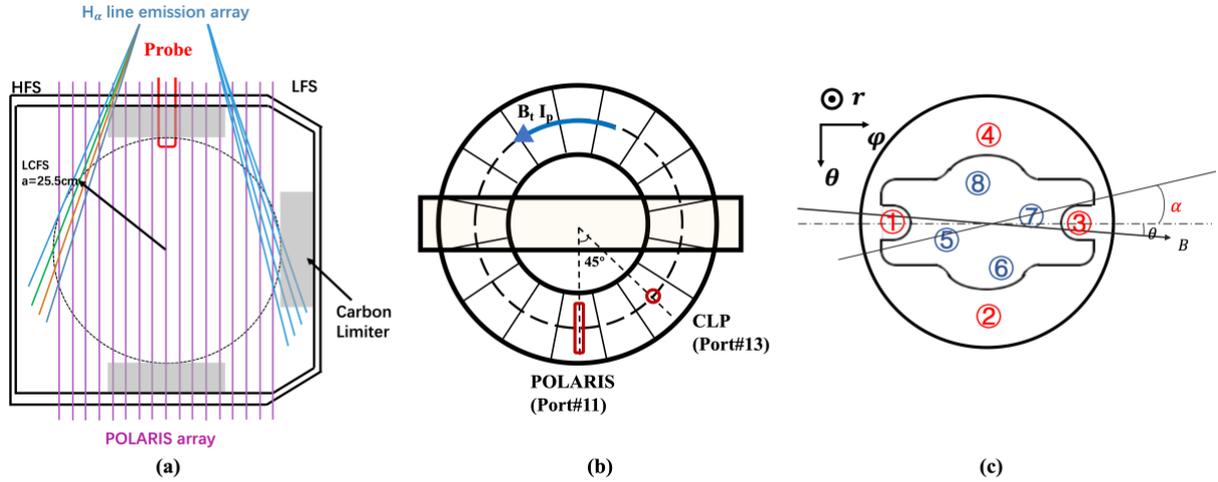

**Figure 1.** Arrangement of the main diagnosis. (a) Cross-section of J-TEXT tokamak, viewing lines of the J-TEXT PDA (in blue color) and POLARIS (in magenta color), respectively. (b) Top view of J-TEXT tokamak, POLARIS and CLP are toroidally separated by 45˚. (c) The configuration of CLP array.

Experimental investigations were undertaken on the J-TEXT tokamak utilizing a limiter configuration in Ohmic hydrogen discharges. Continuous gas puffing was applied throughout the experiments. The ensuing experimental results are based on two distinct discharge conditions, with the corresponding parameters described as follows, unless specailly stated otherwise:

[I] A set of shots with density ramp-up in single shot: plasma current $I_P = 120\ kA$, toroidal magnetic field $B_t = 1.7/1.9/2/2.2\ T$, safety factor $q_a = 4.4/4.9/5/5.5$ at the plasma edge, with the central line-averaged electron density ramping up in the range $2\sim 4.5 \times 10^{19} m^{-3}$, and the Greenwald density limit as $n_G = 5.87 \times 10^{19} m^{-3}$. The CLP remains fixed at $r = 23.5 cm$ in these shots, and a typical plasma traces is shown in Figure 2.

[II] A set of shots with lifting densities shot by shot and constant density in each shot: plasma current $I_P = 120\ kA$, toroidal magnetic field $B_t = 1.7/1.9/2$, safety factor $q_a = 4.4/4.9/5$. The central line-averaged electron density hovers around 1.5 to $3\times 10^{19} m^{-3}$. And the CLP reciprocates to $r = 23.5 cm$ in eash shot, providing the boundary radial profile information under varing electron densities.

For all discharges reported in this paper, the plasma current is sustained as constant, and the boundary safety factor is modified solely by adjusting the toroidal magnetic field, considering that toroidal magnetic field can affect turbulence and transport, but has no effects on density limit.

## 3. Increase of edge turbulence and transport prior to the onset of HFSHD

Previous experiments have shown that the collapse of edge flow shear is a potential trigger for density limit disruption [26]. Interestingly, this collapse of the edge shear flow precedes the actual disruption considerably. In the J-TEXT, the particular behaviour has been observed in the edge region as the line-averaged density approached the density limit. The temporal traces for a typical density ramping discharge are shown in Fig. 2. This discharge has parameters of $I_P = 120\ kA, B_t = 1.7T, q_a = 4.4$. It's notable that the density limit disruption occurs at $t = 600ms$ and the maximum central line-average density is $\bar{n}_{e0} = 4.4 \times 10^{19} m^{-3} = 0.75 n_G$. The central line-averaged electron density keeps ramping up steadily during the constant plasma current ($200ms < t < 600ms$). Fig. 2(e) & (f) present the HFS-LFS edge asymmetries in $H_\alpha$ emission and line-averaged electron density, signifying the emergence of the HFSHD front at $t = 0.48s$ ($\bar{n}_{e0} = 3.4 \times 10^{19} m^{-3}$). Fig. 2(g) compares the visible radiation before (0.44s) and after (0.54s) the arising of HFS-LFS asymmetry. The red dotted line is the estimated last closed flux surface (LCFS). It's obvious that the radiation at the HFS edge is much stronger after the emergence of HFS-LFS asymmetry, and the HFSHD enters into the main plasma region. Fig.2 (h) presents the profiles of 17-channel line-averaged densities, which manifests that the HFSHD is localized at the HFS edge. All these features of HFSHD presented here are highly consistent with our previous results [23]. However, the HFS limiter has been removed in this experiment, and there is no interaction at all between plasma and machine chamber at the HFS edge. This indicates that the impurity and radiation is not important in the formation of HFSHD. But turbulence and transport could be crucial.

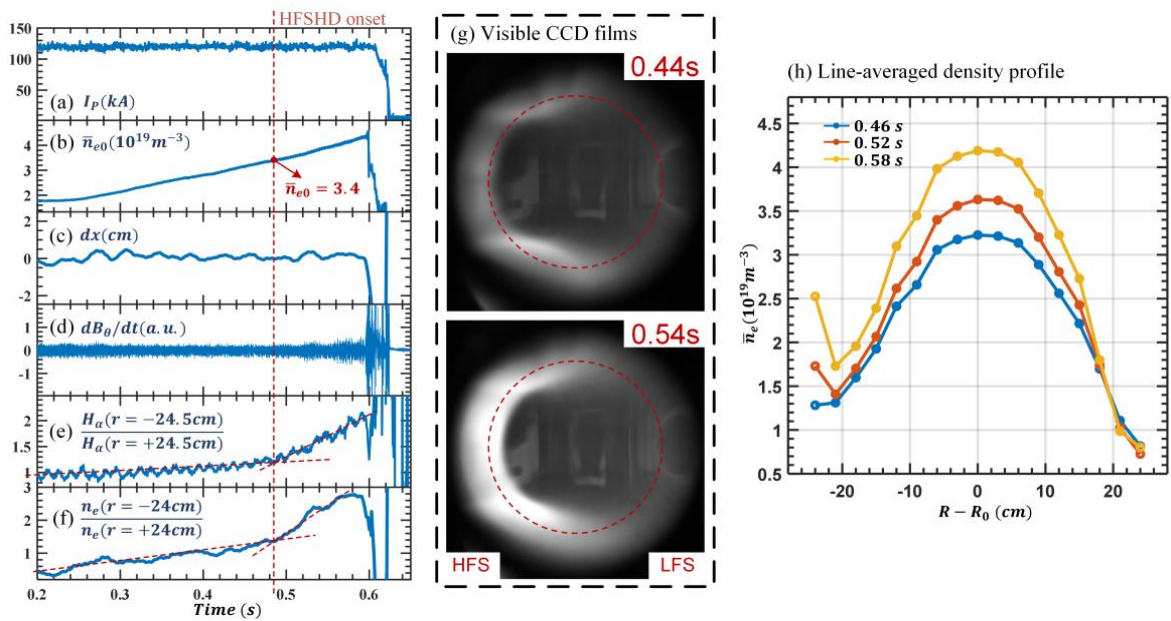

**Figure 2.** A typical density limit disruption discharge. (a) The total plasma current, (b) central line-averaged density measured by FIR polarimeter–interferometer, (c) the plasma horizontal displacement, (d) the edge magnetic coil signal, (e) the $H_\alpha$ emission

asymmetry between the edge channel of HFS and LFS, (f) the density asymmetry between the edge channel of HFS and LFS measured by POLARIS, (g) visible CCD films before (0.44s) and after (0.54s) the formation of HFSHD, and (h) 17-channel line-averaged density profile at three time slices.

Fig. 3 (b-d) displays the temporal evolution of the auto-power spectrum pertaining to (b) the floating potential, (c) the ion saturation current, and (d) the poloidal electric field, as measured by CLP. The shot in Fig. 3 is the same to the one presented in Fig. 2. Meanwhile, Fig. 3(a) presents the temporal evolution of the electron density and total edge particle flux, which is calculated by fluctuations of density and radial velocity.

In details, the auto-power spectrum of the floating potential (Fig. 3(b)) reveals two distinct branches of turbulences, each bearing different characteristic frequencies discernible before $0.4s$. However, the low-frequency ($< 30kHz$) branch of turbulences is unobserved in the auto-power spectrum of ion saturation flow and poloidal electric field, as evidenced in Fig. 3(c) and (d). This implies that the low-frequency turbulence is more indicative of radial electric field fluctuations rather than density fluctuations. Regarding the high-frequency turbulence, it's noteworthy that the characteristic frequency suddenly decreases upon the plasma density exceeding a certain threshold at $0.4s$ (Fig.3(d)), while the amplitude correspondingly increases significantly (Fig.3(c)). Simultaneously, the radial particle flux at the edge experiences a surge of 100% from $0.4s$ to $0.5s$ (Fig.3(a), red lines), while the central line-average electron density merely increases by 10% (Fig. 3(a), blue lines). That clearly indicates that the high-frequency turbulence variations within the high-density region play a crucial role in enhancing particle transport.

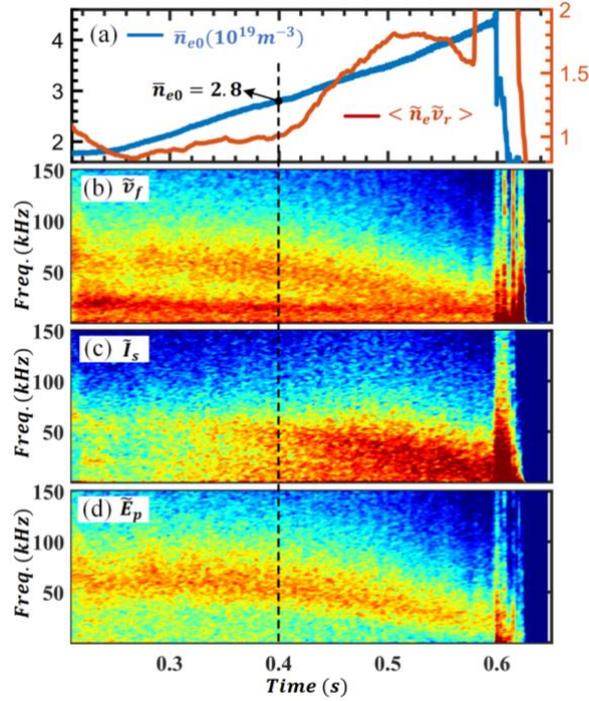

**Figure 3.** Corresponding to figure 2, the time evolution of (a) central line-averaged density (blue lines) and edge particle flux (red lines), auto-power spectrum of (b) floating potential, (c) ion saturation flow, and (d) poloidal electric field.

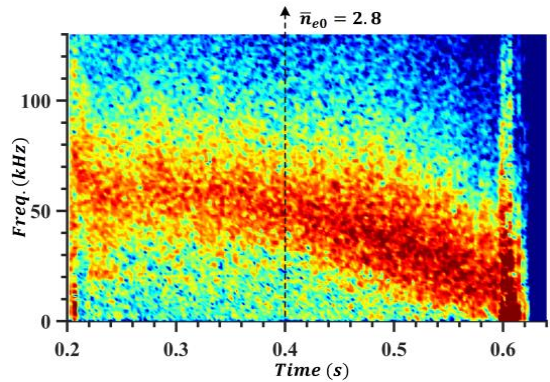

**Figure 4.** Corresponding to figure 3, the auto-power spectrum of approximate turbulent particle flux calculated by $\tilde{E}_\theta \cdot \tilde{I}_s$.

To further study the cause of this turbulent transport enhancement, the auto-power spectrum of the approximate turbulent particle flux calculated by $\tilde{E}_\theta \cdot \tilde{I}_s$ are presented in Fig. 4. From the spectrum of particle flux, it is obvious that the radial particle flux is primarily contributed by the previously discussed broadband turbulence (Fig. 3). Moreover, the low-frequency mode almost does not induce any particle transport, which indicates that it might be GAM.

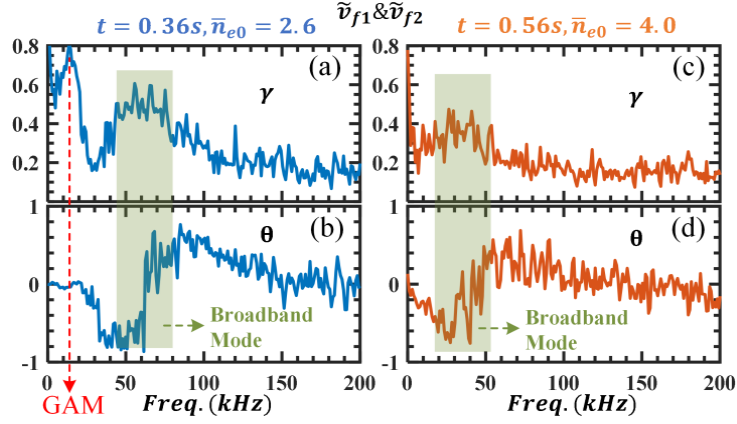

**Figure 5.** Corresponding to figure 3&4, (a) & (c) coherence and (b) & (d) phase difference between two poloidal separated floating potential $\tilde{V}_f$, in different central line-averaged density.

In order to further explore the characteristics of these two different turbulences, Fig. 5 presents the correlation and phase difference between the floating potentials ($\tilde{V}_f$) at two separate poloidal locations. Two cases with different density are compared in Fig. 5. In the case of low density (Fig. 5 (a&b)), the two turbulence branches are markedly distinct. The low-frequency mode exhibits a characteristic frequency of $f \approx 15 kHz$, and a characteristic poloidal wavenumber nearing zero, as suggested by the near-zero phase difference (Fig. 5(b)). These features are consistent to the characteristics of geodesic acoustic modes (GAM) [30], which exhibits toroidal and poloidal symmetry with a finite radial wavenumber. The fact that the low-frequency mode cannot be observed on spectra of ion-saturation flow (Fig. 3(c)) and poloidal electrical field (Fig. 3(d)) further substantiates the characteristics of GAM. Moreover, the theoretical GAM frequency calculated by $f_{GAM} = \sqrt{T_i + T_e/m_i}/2\pi R_0$ is around $16\ kHz$ if we assume that $T_i \approx T_e \approx 35 eV$ (measured by CLP in Fig. 9), which is close to the experimental observation ($15\ kHz$).

Regarding the high-frequency broadband mode, its characteristic frequency remains constant around $60 kHz$ in the low density region ($\bar{n}_{e0} < 2.8 \times 10^{19} m^{-3}$), and decreases to $30 kHz$ preceding the density limit disruption, as indicated by Fig.5 (c). The phase difference between the two potentials depicted in Fig.5 (b) interestingly shows a shift from negative to positive over the peak frequency of the broadband mode. This implies that the wavelength of the mode is close to the distance of the two distributed potential probes ($d = 7mm$). Therefore, it can be deduced that the poloidal wave number of this broadband mode is approximately $k_\theta = 1/d = 1.4 cm^{-1}$. Considering that the edge plasma temperature is around $35 eV$ (showed in Fig. 9), the normalized wave-number stands at about $\rho k_\theta \approx 0.066$, which falls within the range of Ion Temperature Gradient (ITG)/Trapped Electron Mode (TEM). Additionally, it is worth noting that the GAM-like mode disappears in the high-density plasma as shown in Fig.5 (c). This indicates that the enhancement of $I_s$ fluctuations and particle transport in high-density plasmas co-occur with the depression of the GAM-like mode.

Concluding the above experimental observations, after the plasma density surpasses a critical value, several concurrent phenomena are observed at the edge, including the enhancement of broadband turbulence, the suppression of the GAM-like mode, and an increase of radial particle transport. This results suggest that the broadband mode plays a crucial role in boundary transport, and its growth might be related to the suppression of GAM and $E_r$ shear.

In order to investigate the $E_r$ profile evolution as plasma density approaches density limit, we have conducted a series of pulses (/#1079986/#1079987/#1079988) with consistent discharge parameters ($I_P = 120 kA, B_t = 1.7 T$) and incrementally increasing density shot by shot. Then the boundary $E_r$ profile can be obtained by moving the CLP probe during the flat-top phase of density. As demonstrated in Fig. 6, a collapse of the sheared electric field around the last closed flux surface (LCFS) ($r-a=0$) is observed as the line-averaged density escalates from 2 to $2.7 \times 10^{19} m^{-3}$. It's worth noting that the discharge parameters of the three shots in Fig. 6 are identical to the shot in Fig. 3&4&5. Thus, they should share the same critical density threshold for the enhancement of turbulence and transport. By synthesising the results from Fig. 3 and Fig. 6, we can infer that the collapse of the sheared radial electric field in the edge region is followed by a sudden increase in the intensity of the ion saturation flow perturbation and the edge particle flux.

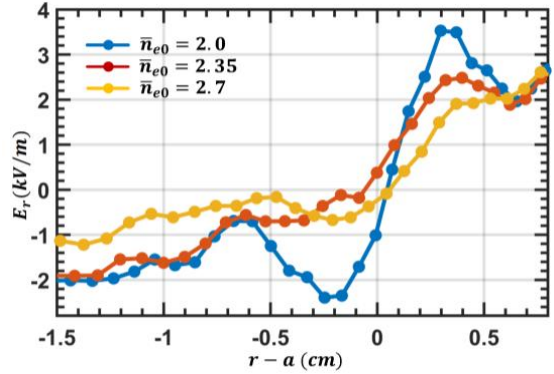

**Figure 6.** Profiles of edge radial electric field $E_r$. These profiles are obtained while keeping the plasma density constant in a single shot. The blue, red and yellow lines represent three line-averaged densities $\bar{n}_{e0} = 2/2.35/2.7 \times 10^{19} m^{-3}$ respectively.

As illustrated in previous publications [3, 23, 36], the MARFE or HFSHD is the direct cause of MHD instability and major disruption. And the enhancement of turbulence and transport could be responsible for edge cooling and the emergence of MARFE or HFSHD. The rest of this paper will primarily explore the correlation between the turbulent transport and HFSHD. As stated in our previous publication [23], the occurrence of the HFSHD front can be identified by the ratio of the line-averaged density at the edge between HFS and LFS. As demonstrated in Fig. 2, when the central line-average density exceeds $\bar{n}_{e0} > 3.4 \times 10^{19} m^{-3}$, the HFS line-average density increases rapidly and the asymmetry develops significantly. This indicates that a local high-density plasma region is formed in the HFS edge as the plasma density exceeds a critical value ($n_{crit} = 3.4 \times 10^{19} m^{-3}$ in this discharge). Such observations are representive of the characteristic features of the HFSHD front phenomenon, as described in [19-20, 23].

To summarize the sequential phenomena as the plasma density approaches to the density limit, a collapse of the sheared electric field is first observed, followed by an abrupt increase in electron density fluctuations and edge particle flux, and later on the HFSHD front appears. Based on the sequence of these events, it can be speculated that the collapse of $E_r$ shear and subsequent increase in turbulent transport is the primary trigger for density limit disruption on tokamaks. Thermal-radiation instabilities like MARFE and HFSHD, as the direct triggers for MHD and major disruption, are simply the outcomes of increased transport.

## 4. Effects of edge safety factor on boundary turbulent transport and HFSHD formation

To further verify the correlation between the increase in turbulent transport and the onset of the HFSHD front, we investigated the development of the boundary turbulence and HFSHD formation in plasmas with varied $q_a$, considering that $q_a$ can affect turbulence and transport, but is weakly related to density limit [3].

In Fig. 7, we compare the traces of radial electric field $E_r$ shear rate around the LCFS, edge particle flux and HFS-LFS density asymmetry as a function of line-average density for the shot shown in Fig. 3 (where $q_a = 4.4$) and another shot with a higher $q_a$ value of 5. The high $q_a$ discharge has parameters: $I_P = 120kA, B_t = 2, q_a = 5$. The evolution of particle flux and HFS-LFS density asymmetry are obtained from the discharges with increasing densities in single shot (set [I]), and the $E_r$ shear rate is acquired from the experiments with lifting densities shot by shot (set [II]).

It is noteworthy that both shots undergo a series of events including the collapse of $E_r$ shear, a surge in boundary particle flux, and the emergence of the HFSHD front. Interestingly, the density threshold for the onset of HFSHD front is significantly lower in the high $q_a$ dsicharge compared to the low $q_a$ one, as showed in Fig. 7 (c). This feature of HFSHD front is basicly different to MARFE, the denstiy threshold of which is found to be independent with $B_t$ or $q_a$ [37]. In addition, the critical densities for the collapse of $E_r$ shear and subsequent enhancement of particle transport also demonstrate lower values in the high $q_a$ shot relative to the low $q_a$ shot.

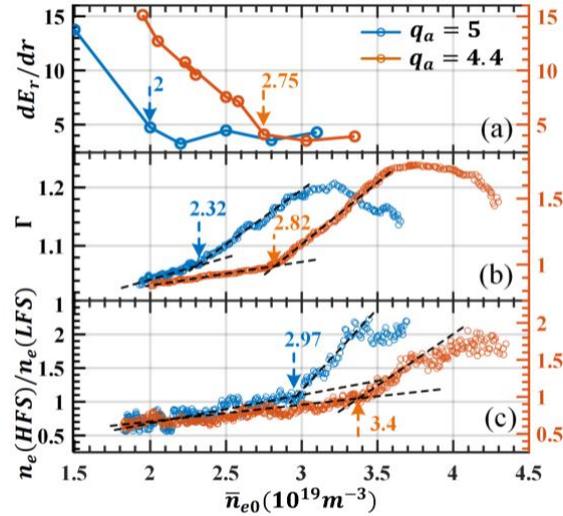

**Figure 7.** Traces of (a) radial electric field ($E_r$) shear rate, (b) edge particle flux and (c) HFS-LFS density asymmetry against line-average density for $q_a = 5$ (blue lines) and $q_a = 4.4$ (red lines).

In fact, the behaviors of turbulence and transport presented herein can elucidate why the density threshold of HFSHD front is inversely proportional to $q_a$. As reported in Ref. [23], a critical value of edge collisionality is the trigger for the onset of HFSHD front. The enhancement of edge turbulent transport will accelerate the edge cooling and increases edge collisionality. Therefore, in a high $q_a$ discharge, the lower density threshold of the HFSHD front is a consequence of the lower critical density value associated with the increase in turbulent transport.

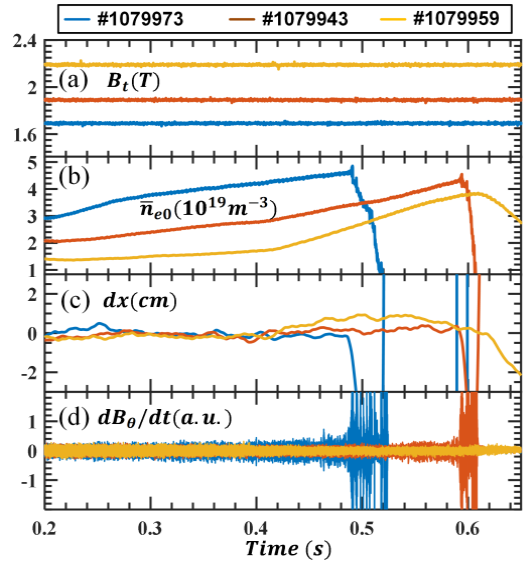

**Figure 8.** Time evolution of the main parameters for three typical discharges (#1079973/#1079943/#1079959) with ramping density in J-TEXT. (a) toroidal field, (b) central line-average electron density measured by POLARIS, (c) horizontal plasma displacement and (d) the edge Mirnov coil signal.

The experimental observations detailed above suggest that the edge turbulent transport plays an important role in the emergence of the HFSHD front. At the same time, the HFSHD front also impacts the evolution of boundary parameters. Discharges in the following were conducted with a constant plasma current of $I_P = 120kA$ and a varying toroidal field $B_t = 1.7/1.9/2.2\ T$, in order to scan the safe factor while maintaining the Greenwald density limit. Fig. 8 shows the main traces for the three typical discharges (#1079973 / #1079943 / #1079959) with ramping density in J-TEXT. In these shots, the CLP was sustained at $r = 23.5cm$, to obtain the evolution of edge parameters along with the density increasing. The plasma density is increased by continuous gas-puffing in each shot, as showed in Fig. 8(b). Also, shots #1079973 and #1079943 disrupt at similar density, caused by density limit. This result confirm that density limit is weakly related to $B_t$ or $q_a$. The three plasmas in Fig. 8 have similar position, as indicated by horizontal displacement in Fig. 8 (c). In addition, the poloidal magnetic field fluctuations measured by Mirnov coil in Fig. 8 (d) indicate that there is no macro-MHD activities before density limit disruption. And the magnetic fluctuations in higher $B_t$ plasma is significantly lower than that in lower $B_t$ plasma, suggesting that higher $B_t$ or $q_a$ can stabilize the MHD instabilities.

Corresponding to the three shots in Fig. 8, Fig. 9 shows the HFS-LFS density asymmetry and edge parameters measured by CLP at $r = 23.5cm$ against the central line-averaged density. The rapid rise in density asymmetry observed in Fig. 9(a) is recognized as a characteristic feature of the emergence of the HFSHD front. Notably, the critical density threshold for the appearance of the HFSHD front appears to be inversely proportional to $q_a$. These rerults are consistent with the findings in Ref. [23].

Fig. 9 (b) and (c) present the floating potential and electron temperature at $r = 23.5cm$ respectively. Evidently, both of them persistently decrease prior to the appearance of the HFSHD front. To some extent, the evolution of floating potential at $r = 23.5cm$ can represent the trace of $E_r$ shear rate around LCFS, considering that the floating potential is consistently near to 0 at LCFS ($r = 25.5cm$) in Ohmic plasmas. Consequently, this infers a continuous reduction in the $E_r$ shear rate preceding the onset of the HFSHD front. Importantly, the changes in the boundary $E_r$ shear rate and $T_e$ with respect to plasma density are notably influenced by the edge safety factor $q_a$. The decrement of $V_f$ and $T_e$ at the edge is discernibly more premature in low $q_a$ plasmas compared to high $q_a$ discharges. In other words, the values of boundary $E_r$ shear rate and $T_e$ are lower in discharges with a higher $q_a$, given that the plasma densities are the same.

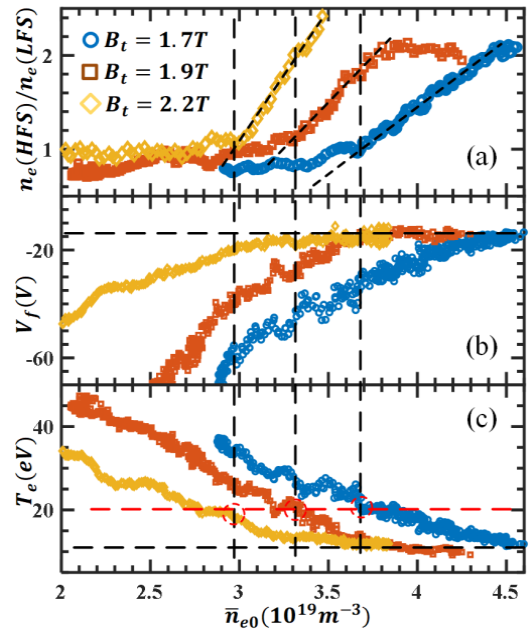

**Figure 9.** Traces of HFS-LFS density asymmetry measured by POLARIS (a) and edge parameters measured by CLP against line-average density. (b) Floating potential and (c) electron temperature for different $q_a$.

Subsequent to the emergence of the HFSHD front, both the edge $V_f$ and $T_e$ tend to saturate at certain values. And these critical density thresholds at which saturation occurs are inversely proportional to $q_a$, which mirrors the behavior of density threshold of the HFSHD front. Interestingly, the final saturation values for both the floating potential and electron temperature appear to be independent of $q_a$. This implies that the plasma parameters within the radial region affected by HFSHD front are quite stable, regardless of the $q_a$ value. As pointed out in our previous publication [23], the HFSHD front stems from the HFS SOL region, and expands radially and poloidally. A fully developed HFSHD front can extend to a radial location of $r = 20cm$, which is significantly deeper than the location of CLP ($r = 23.5cm$). As such, the CLP location is significantly impacted by the HFSHD front once the front is adequately detected by POLARIS. Moreover, according to the one-dimensional flux tube model at the plasma edge (Fig. 7 in Ref. [23]), the LFS end serves as the heat source of the flux tube, under the assumption that the radial heat transport is dominant by the ballooning mode turbulence. Meanwhile, the HFS end, where the cold-dense HD front resides, acts as the heat sink. Therefore, there should have a parallel heat flow from the LFS to HFS end, ensuring the maintenance of thermal equilibrium. The CLP measures the top of plasma, situated in the middle of the flux tube. The observed results indicate that the plasma parameters in the mid-section of the flux tube where the HFSHD front is located, are stable. This further confirms that the HFSHD front is quite stable in contrast to MARFE. Besides, as shown by Fig. 9(c), the electron temperature in the flux tube affected by HFSHD front appears to stabilize around $\sim 10eV$. This is consistent with the peak of the radiation cooling rate observed for carbon impurity, as discussed in Ref. [38].

In addition, it's worth noting that the HFSHD front consistently occurs when the edge electron temperature drops to around $\sim 20eV$, irrespective of the specific $q_a$ value in the discharge. This suggests that the edge temperature plays a critical role in the formation of HFSHD front. The evolution of edge electron temperature further corroborates the link between the edge turbulent transport and the onset of HFSHD front. The edge temperature is mainly determined by radial transport in Ohmic plasmas, given that the heating source is centralized at the center. The fact that the drop of edge temperature is later in high $q_a$ discharges than the low $q_a$ ones, supports the above hypothesis that enhancement of turbulent transport plays a crucial role in creating the conditions for the HFSHD front occurrence, likely by leading to increased edge cooling and higher collisionality at the plasma edge.

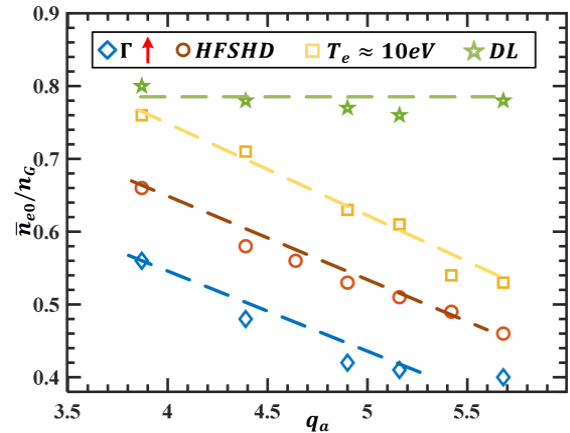

**Figure 10.** Statistics of density thresholds against safety factor $q_a$ for the rapid increasing of partical flux (blue diamond), HFSHD front emergence (red circles), boundary $T_e$ drops to $10eV$ (yellow squares), and the maximum density (green pentagram).

Figure 10 shows the statistics of density thresholds for all the above physical phenomena against boundary safety factor $q_a$. The series of observations documented provides valuable insights into the sequence of events that occur during plasma density ramp-ups under different $q_a$ in tokamak experiments. The unfolding of events – from the collapse of shear flow, to the swift enhancement of boundary turbulent transport, the emergence of the HFSHD front, the saturation of boundary temperature, and ultimately, the density limit disruption – is strikingly consistent, suggesting a potentially universal mechanism underlying density limit disruption in tokamak plasmas.

Futhermore, the data show that the density thresholds for all the above physical phenomena bear an inverse relationship to $q_a$. This suggests that higher $q_a$ values prompt the edge cooling and HFSHD front to manifest earlier in the plasma density ramp-up process. Nevertheless, the point of density limit disruption itself doesn't display a significant shift for varying $q_a$. This could potentially be attributed to the premature onset of edge cooling and HFSHD front, which act as catalysts for density limit disruption, while concurrently, a high $q_a$ works to stabilize MHD instability.

## 5. Summary and Discussion

In this paper, we report the discovery and analysis of boundary turbulence behaviour and the HFSHD front as the density approaches density limit on the J-TEXT tokamak. It's noteworthy that the HFSHD front still occurs even though the HFS limiter has been removed. This highlights that the HFSHD front is a consequence of transport, rather than being tied to impurity radiation, which is associated with plasma-wall interaction. The experiments were carried out by Ohmic heating and the plasma density was increased by continuous gas puffing without the use of auxiliary system or external drives. The following is a summary of the experimental findings:

(a) Two different branches of turbulences are observed by Langmuir probes at $\rho = 0.92a$. The low-frequency ($f \approx 15kHz$) branch of turbulence is identified as Geodesic Acoustic Mode (GAM), characterized by its dominance in potential fluctuations, zero poloidal wavenumber ($k_\theta \approx 0$) and lack of contribution to radial particle transport. The high-frequency ($50 \sim 80kHz$) turbulence is a broad-band mode that can be identified in the auto-power spectrum of the ion saturation flow and poloidal electric field. The edge particle flux is found to be mainly contributed by the high-frequency trubulence.

(b) After the plasma density exceeds a critical value, there is a sudden amplitude increase in high-frequency turbulence, along with a decrease in its frequency. Concurrently, the GAM is suppressed and the radial particle flux increases. In addition, a collapse of $E_r$ shear around the LCFS is observed at a lower density than threshold for turbulent transport enhancement. Furthermore, the appearance of the HFSHD front follows the increase of boundary turbulence.

(c) The edge floating potential and electron temperature measured by CLP consistently decrease after the increase of turbulent transport. Interestingly, they appear to stabilize at a certain value after the onset of HFSHD front. In various $q_a$ discharges, the HFSHD front always accurs when the edge $T_e$ drops to around $\sim 20eV$, even though the density threshold is apparently different.

Based on the above observations, we can outline a series of boundary events leading to density limit disruption. By sequence, they are the collapse of $E_r$ shear, enhancement of turbulent transport, suppression of GAM, $T_e$ decrease and the appearance of HFSHD front. This sequence indicates that the collapse of $E_r$ shear and subsequent enhancement of turbulent tranport is the primary cause for density limit disruption in tokamak. The HFSHD front, which is the direct cause for MHD and major disruption, is just the result of enhancement of turbulent transport.

Importantly, the pattern of sequential events is consistent in various density climbing discharges. And the density thresholds for these phenomena are all inversely proportional to $q_a$. These results further emphasize the crucial role that edge turbulent transport plays in density limit disruption. In higher $q_a$ discharge, the lower density threshold for edge cooling and the HFSHD front can be attributed to the earlier collapse of $E_r$ shear and an increase in turbulent transport during the ramp-up of plasma density. Additionally, the impact of $q_a$ on the density threshold for the collapse of $E_r$ shear can be elucidated through the theory proposed by Hajjar and Diamond [39]. According to their theory, as the plasma response passes from adiabatic ($\alpha > 1$) to hydrodynamic ($\alpha < 1$), the edge zonal shear layer collapses and turbulence is enhanced. The key variable $\alpha$, the adiabaticity parameter, equals $k_z^2 v_{th}^2/(v_{ei}|\omega|)$. Here, $k_z$ represents the parallel wavenumber, which can be estimated to be $\sim 1/qR$. As a result, $\alpha$ is inversely dependent on $q_a$, suggesting a lower density threhold for collapse of edge shear layer in a higher $q_a$ discharge.

Interestingly, the density limit itself exhibits a very weak dependence on $q_a$. This could potentially be interpreted as the cumulative effects of $q_a$ on micro-turbulence and macro-MHD. On one hand, higher the $q_a$ could lead to a lower the density threshold for the collapse of $E_r$ shear and resultant HFSHD front, which is detrimental to plasma confinement. On the other hand, higher $q_a$ values tend to stabilize the MHD instability. Taken together, from a qualitative point of view, different $q_a$ values do not significantly impact the final value of density limit disruption. It is important to note that the above speculations are qualitative rather than quantitative. We will further conduct simulation studies in the future to demonstrate the feasibility of these speculations.

**Acknowledgements**: This work is supported by the National MCF Energy R&D Program under Grant No. 2018YFE0310300 and the National Natural Science Foundation of China under Grant No. 11905080 and 51821005. This work has been also part-funded by the EPSRC Energy Programme [grant number EP/W006839/1]